\documentclass[a4paper,10pt,twoside]{just}

\usepackage{multicol}
\usepackage{graphicx}
\usepackage{booktabs}
\usepackage{amssymb,bm,mathrsfs,bbm,amscd}
\usepackage[tbtags]{amsmath}
\usepackage{lastpage}
\usepackage{CJK}
\usepackage{epstopdf}

\begin{document}
\begin{CJK*}{GBK}{song}

\fancyhead[c]{\small  10th International Workshop on $e^+e^-$ collisions from $\phi$ to $\psi$ (PhiPsi15)}
 \fancyfoot[C]{\small PhiPsi15-\thepage}

\footnotetext[0]{Received 20 Nov. 2015}

\title{Hadronic contribution from light by light processes in (g-2) of muon in nonlocal quark model.\thanks{The work is supported by Russian Science Foundation
		Grant (RSCF 15-12-10009) }}

\author{%
      A.~S.~Zhevlakov$^{1;1)}$\email{zhevlakov@phys.tsu.ru}
\quad A.~E.~Dorokhov$^{2;2)}$\email{dorokhov@theor.jinr.ru} 
\quad A.~E.~Radzhabov$^{3;3)}$\email{aradzh@icc.ru} 
}
\maketitle

\address{%
$^1$ Department of Physics, Tomsk State University, Lenin ave. 36, 634050 Tomsk, Russia\\
$^2$   Bogoliubov Laboratory of Theoretical Physics, JINR, 141980 Dubna, Russia\\
$^3$ 3 Institute for System Dynamics and Control Theory SB RAS, 664033 Irkutsk, Russia\\
}

\begin{abstract}
The hadronic corrections to the muon anomalous magnetic moment $a_{\mu}$, due to the full
gauge-invariant set of diagrams with dynamical quark loop and intermediate pseudoscalar and scalar states light-by-light scattering insertions, are calculated in the framework of the
nonlocal chiral quark model.
These diagrams correspond to all hadronic light-by-light scattering contributions to $a_{\mu}$ in the leading order of the $1/N_{c}$ expansion in quark model. 
The result for the quark loop contribution
is $a_{\mu}^{\mathrm{HLbL,Loop}}=\left(  11.0\pm0.9\right)  \cdot10^{-10},$
and the total result is $a_{\mu}^{\mathrm{HLbL,N\chi QM}}=\left(
16.8\pm1.2\right)  \cdot10^{-10}$.

\end{abstract}



\begin{multicols}{2}

\section{Introduction}

The anomalous magnetic moment (AMM) of lepton and contribution of light by light (LbL) processes
has a long history of investigation. 
After latest experiment on measurement of AMM of muon in 
BNL E821 \cite{Bennett:2006fi} the interest to this topic are return. 
Two new experiments on measurement of AMM of muon are under construction in Fermilab \cite{Venanzoni:2012sq} and J-PARC \cite{Saito:2012zz,Shwartz_talk}. New precision data are demand more accurate calculations.

The most problematic part of the calculation AMM of muon is segment associated with the strong interaction because most of this contribution is in nonperturbative low energy region. 
This contribution consist of hadron vacuum polarization (HVP) part (leading in $\alpha$) and LbL scattering through the nonperturbative QCD vacuum (sub-leading in $\alpha$). The HVP contribution can be extracted from the experimental data but the contribution of LbL scattering needs to be modeled. 
  
What degrees of freedom (DoF)
are relevant for modeling of strong interaction at low energy: 
mesons or quarks (and gluons)? 
This question is connected with confinement problem and is one of the most important tasks in physics of strong interaction.
One can separate two different approaches for description of LbL processes.
In first one the only mesonic DoF are used. 
The second one starts from quark Lagrangian and have mesonic DoF as a bound states.  

\section{Model}

The LbL contribution to AMM of muon is in low energy region where the perturbative methods of QCD are not applicable. 

Nonlocal quark model  N$\chi$QM is nonlinear realization of Nambu-Jona-Lasinio model. Nonlocality can be motivated by instanton liquid model. The model is formulated in terms of quark degrees of freedom and bound states corresponds to mesons. 
The circumscribing of model is made in works \cite{Scarpettini:2003fj,Anikin:2000rq} and here we give brief description model properties that is needed for calculation of AMM.

\subsection{Lagrangian}
The Lagrangian of the SU(3) nonlocal chiral quark model with
the SU(3)$\times$SU(3) symmetry has the form

\begin{align}
&\mathcal{L}   =\bar{q}(x)(i\hat{\partial}-m_{c})q(x)+\frac{G}{2}[J_{S}%
^{a}(x)J_{S}^{a}(x)+J_{PS}^{a}(x)J_{PS}^{a}(x)]\nonumber\\
&-\frac{H}{4}T_{abc}\Big[J_{S}^{a}(x)J_{S}^{b}(x)J_{S}^{c}(x)-3J_{S}%
^{a}(x)J_{PS}^{b}(x)J_{PS}^{c}(x)\Big],\label{Model}
\end{align}
where $q\left( x\right) $ are the quark fields, $m_{c}$ $\left(
m_{u}=m_{d}\neq m_{s}\right) $ is the diagonal matrix of the quark current
masses, $G$ and $H$ are the four- and six-quark coupling constants. Second
line in the Lagrangian represents the Kobayashi--Maskawa--t`Hooft
determinant vertex with the structural constant

\begin{equation}
T_{abc}=\frac{1}{6}\epsilon _{ijk}\epsilon _{mnl}(\lambda _{a})_{im}(\lambda
_{b})_{jn}(\lambda _{c})_{kl},
\end{equation}%
where $\lambda _{a}$ are the Gell-Mann matrices for $a=1,..,8$ and $\lambda
_{0}=\sqrt{2/3}I$.

The nonlocal structure of the model is introduced via the nonlocal quark
currents
\begin{equation}
J_{M}^{a}(x)=\int d^{4}x_{1}d^{4}x_{2}\,f(x_{1})f(x_{2})\,\bar{q}%
(x-x_{1})\,\Gamma _{M}^{a}q(x+x_{2}),  \label{JaM}
\end{equation}%
where $M=S$ for the scalar and $M=PS$ for the pseudoscalar channels, $\Gamma
_{{S}}^{a}=\lambda ^{a}$, $\Gamma _{{PS}}^{a}=i\gamma ^{5}\lambda ^{a}$ and $%
f(x)$ is a form factor with the nonlocality parameter $\Lambda $ reflecting
the nonlocal properties of the QCD vacuum.

The model can be bosonized using the stationary phase
approximation which leads to the system of gap equations for the dynamical
quark masses $m_{d,i}$%
\begin{equation}
m_{d,i}+GS_{i}+\frac{H}{2}S_{j}S_{k}=0,  \label{GapEqs}
\end{equation}%
with $i=u,d,s$ and $j,k\neq i,$ and $S_{i}$ is the quark loop integral
\begin{equation*}
S_{i}=-8N_{c}\int \frac{d_{E}^{4}k}{(2\pi )^{4}}\frac{%
	f^{2}(k^{2})m_{i}(k^{2})}{D_{i}(k^{2})},
\end{equation*}%
where $m_{i}(k^{2})=m_{c,i}+m_{d,i}f^{2}(k^{2})$, $%
D_{i}(k^{2})=k^{2}+m_{i}^{2}(k^{2})$ is the dynamical quark propagator
obtained by solving the Dyson-Schwinger  equation, $f(k^{2})$ is the nonlocal form factor in the momentum representation. For calculation was used two different form-factors: Gaussian form 	
\begin{equation}
f(p^2)=\exp\left(-\frac{p^2}{2\Lambda^2}\right)
\end{equation}
monopole form
\begin{equation}
f(p^2)=\left(1+\frac{p^2}{\Lambda^2}\right)^{-1}
\end{equation}
where $\Lambda$ is cutoff parameter. The model have five parameters which can fitted on physical observables. In order to investigate the sensitivity of model to the changing of model parameters the dynamical mass of light quark is varying between 200-350 MeV with corresponding refit of other parameters. This region corresponds to the more or less physical range of dynamical quark mass.
\begin{figure*}
	\begin{center}
		\begin{tabular*}{\textwidth}{@{}cccccccccc@{}}
			\qquad&\raisebox{-0.5\height}{\resizebox{!}{0.062\textwidth}{\includegraphics{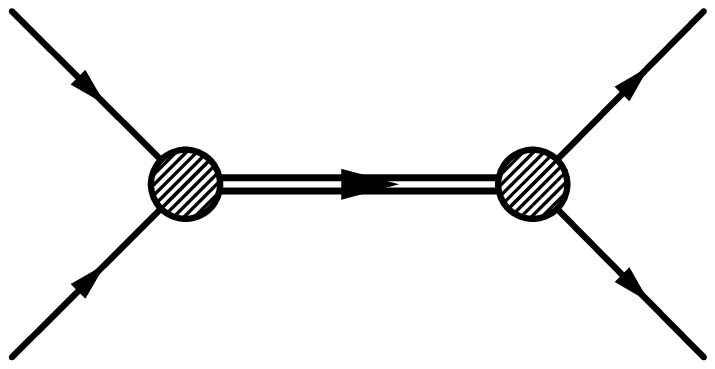}}}&\raisebox{-1\height}{\textbf{=}}&
			\raisebox{-0.5\height}{\resizebox{!}{0.062\textwidth}{\includegraphics{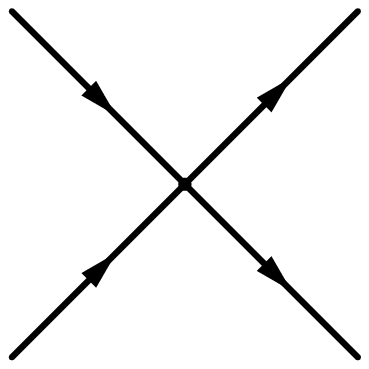}}}&\raisebox{-1\height}{\textbf{+}}&
			\raisebox{-0.5\height}{\resizebox{!}{0.062\textwidth}{\includegraphics{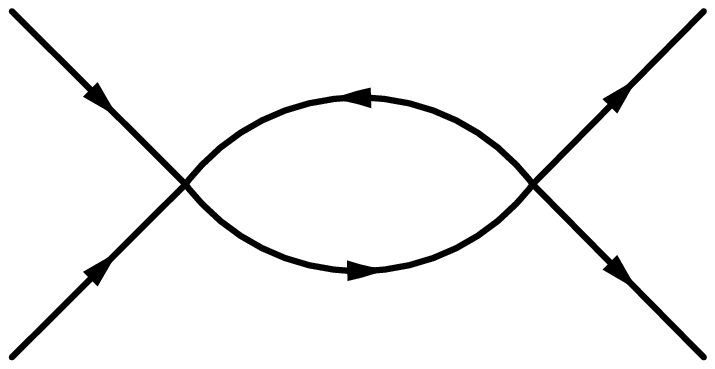}}}&\raisebox{-1\height}{\textbf{+}}&
			\raisebox{-0.5\height}{\resizebox{!}{0.052\textwidth}{\includegraphics{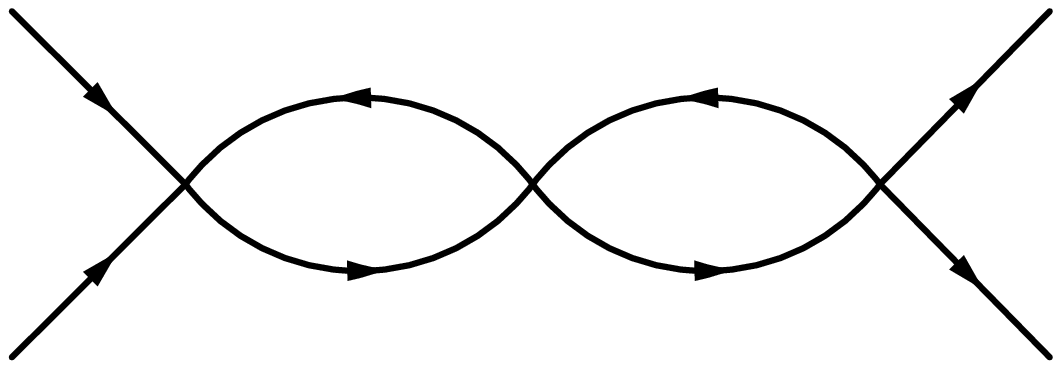}}}&\raisebox{-1\height}{\textbf{+}}&\raisebox{-1\height}{\textbf{...}}\\
			&&&&&&&&&\\
			\qquad\qquad\qquad\qquad&\raisebox{-0.52\height}{\resizebox{!}{0.06\textwidth}{\includegraphics{Qmeson}}}&\raisebox{-1\height}{\textbf{=}}&
			\raisebox{-0.5\height}{\resizebox{!}{0.062\textwidth}{\includegraphics{4t}}}&\raisebox{-1\height}{\textbf{+}}&
			\raisebox{-0.5\height}{\resizebox{!}{0.062\textwidth}{\includegraphics{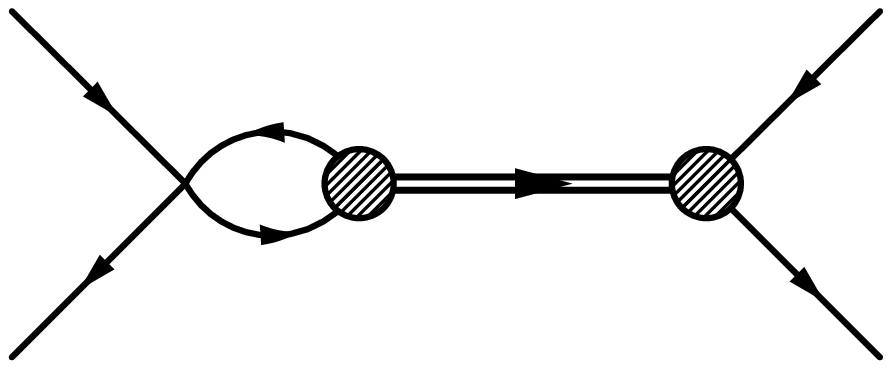}}}&&
		\end{tabular*} 
		\figcaption{ \label{fig:prop}The set of diagrams with four-quarks interaction vertex can be represented by pure four-quark vertex and sum of diagram that will associated with meson exchange.}\end{center}
\end{figure*}

\subsection{Meson propagator}
The quark-meson vertex functions and the meson masses can be found from the
solution of Bethe-Salpeter equation Fig.~\ref{fig:prop}. For the separable interaction \cite{Anikin:2000rq} the quark-antiquark scattering matrix in each ($PS$ or $S$)
channels becomes
\begin{align}
& \mathbf{T}=\hat{\mathbf{T}}(p^{2})\delta ^{4}\left(
p_{1}+p_{2}-(p_{3}+p_{4})\right) \prod\limits_{i=1}^{4}f(p_{i}^{2}),  \notag
\\
& \hat{\mathbf{T}}(p^{2})=i\gamma _{5}\lambda _{k}\left( \frac{1}{-\mathbf{G}%
	^{-1}+\mathbf{\Pi }(p^{2})}\right) _{kl}i\gamma _{5}\lambda _{l},
\end{align}%
where $p_{i}$ are the momenta of external quark lines, $\mathbf{G}$ and $%
\mathbf{\Pi }(p^{2})$ are the corresponding matrices of the four-quark
coupling constants and the polarization operators of mesons ($%
p=p_{1}+p_{2}=p_{3}+p_{4}$). The meson masses can be found from the zeros of
determinant $\mathrm{det}(\mathbf{G}^{-1}-\mathbf{\Pi }(-M^{2}))=0.$ The $%
\hat{\mathbf{T}}$-matrix for the system of mesons in each neutral channel
can be expressed as
\begin{equation}
\hat{\mathbf{T}}_{ch}(P^{2})=\sum_{M_{ch}}\frac{\overline{V}%
	_{M_{ch}}(P^{2})\otimes V_{M_{ch}}(P^{2})}{-(P^{2}+\mathrm{M}_{{M_{ch}}}^{2})},  \label{Tch}
\end{equation}%
where $\mathrm{M}_{M}$ are the meson masses, $V_{M}(P^{2})$ are the vertex
functions $\left( \overline{V}_{M}(p^{2})=\gamma ^{0}V_{M}^{\dag
}(P^{2})\gamma ^{0}\right) $. The sum in (\ref{Tch}) is over full set of
light mesons: $(M_{PS}={\pi ^{0},\eta ,\eta ^{\prime }})$ in the
pseudoscalar channel and $(M_{S}={a_{0}(980),f_{0}(980),\sigma })$ in the
scalar one.

Details about vertex of interaction of mesons with quarks, meson propagator, etc. mixing 
can be found in 
\cite{Scarpettini:2003fj,Dorokhov:2011zf,Dorokhov:2012qa}. 


\subsection{Interaction with external photons}

The bext step for description LbL processes is to introduce in the nonlocal chiral Lagrangian (\ref{Model}) the gauge-invariant interaction with an external
photon field $A_{\mu}(z)$  by Schwinger factor (\ref{SchwPhF}). In result we obtain infinite series of vertexes quark-antiquark interactions with photons.  

	\begin{center}
		\centerline{\begin{tabular*}{\columnwidth}{@{\extracolsep{\fill}}cccc}
				\resizebox{!}{0.10\textheight}{\includegraphics{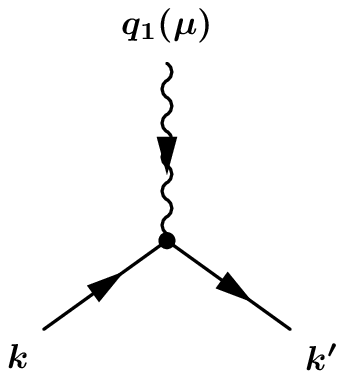}}
				& \resizebox{!}{0.10\textheight}{\includegraphics{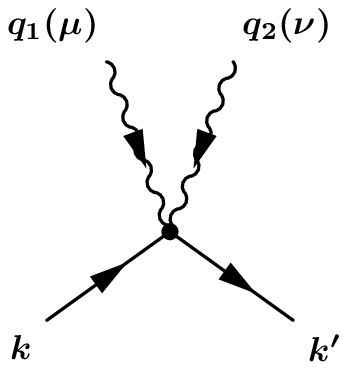}}\\
				(a) & (b) \\\\ 
				\resizebox{!}{0.10\textheight}{\includegraphics{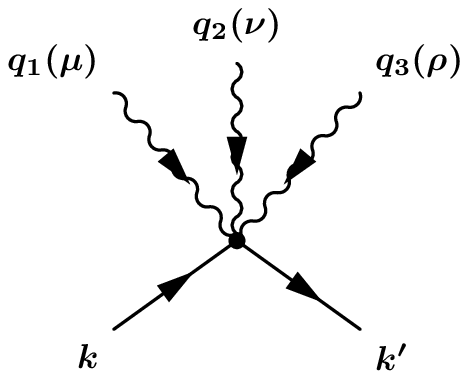}}
				& \resizebox{!}{0.10\textheight}{\includegraphics{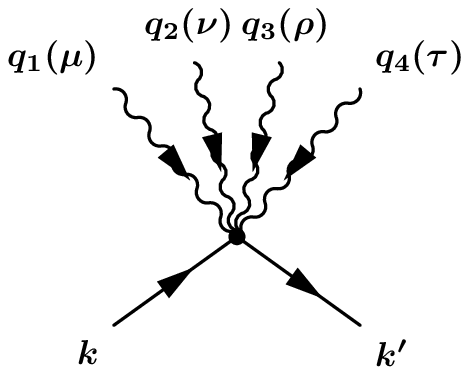}}
				\\
				(c) & (d)
			\end{tabular*}}
			
			\figcaption{\label{fig:VerticesPhot-n}The quark-photon vertex $\mathrm{\Gamma}_{\mu}^{\left(  1\right)
				}\left(  q\right)  $, the quark-two-photon vertex $\mathrm{\Gamma}_{\mu\nu
			}^{(2)}\left(  q_{1},q_{2}\right)  $, the quark-three-photon vertex
			$\Gamma_{\mu\nu\rho}^{(3)}(q_{1},q_{2},q_{3}),$ and the quark-four-photon
			vertex $\Gamma_{\mu\nu\rho\tau}^{(4)}(q_{1},q_{2},q_{3},q_{4})$ . }%
	\end{center}
	
\begin{equation}
q\left(  y\right)  \rightarrow Q\left(  x,y\right)  =\mathcal{P}\exp\left\{
i\int_{x}^{y}dz^{\mu}A_{\mu}\left(  z\right)  \right\}  q\left(  y\right)  .
\label{SchwPhF}%
\end{equation}
 The 
 scheme, based on the rules that the
 derivative of the contour integral does not depend on the path shape
 \begin{align}
 \frac{\partial}{\partial y^{\mu }}\int\limits_{x}^{y}dz^{\nu }\
 F_{\nu}(z)=F_{\mu }(y),\quad \delta^{(4)}\left( x-y\right)
 \int\limits_{x}^{y}dz^{\nu}\ F_{\nu }(z)=0,
 \nonumber
 \end{align}
 was suggested in \cite{Mandelstam:1962mi}
 and applied to nonlocal models in \cite{Terning:1991yt}.
The actual form of 
the vertexes shown in Fig.~\ref{fig:VerticesPhot-n} 
can be found in \cite{Dorokhov:2015psa}.

\subsection{Box diagram}

In effective quark model under consideration there are 
two different parts which corresponds to contact contribution or contribution with intermediate meson. 

Using quark-antiquark interactions vertexes with one, two, three of four photons, see Fig.~\ref{fig:VerticesPhot-n}, we can build five types of diagrams.  

\end{multicols}{2}
\ruleup
\vspace*{6mm}
\begin{center}
	\centerline{\includegraphics[width=0.65\textwidth]{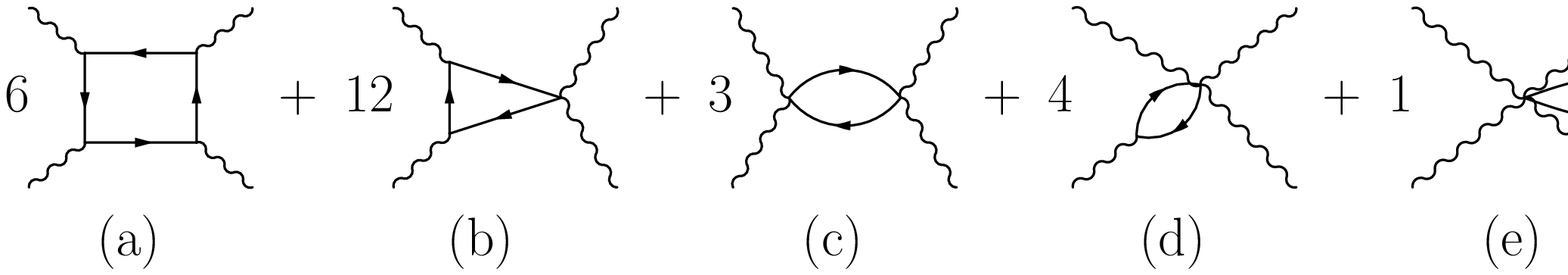}} \vspace*{8pt}%
	\figcaption{	\label{Fig: BoxCont}The box diagram and the diagrams with nonlocal multiphoton
		interaction vertices that give the contributions to $\mathrm{\Pi}_{\mu
			\nu\lambda\rho}(q_{1},q_{2},q_{3})$. The numbers in front of the diagrams are
		the combinatoric factors. }%
\end{center}
\ruledown
\vspace*{4mm}

\begin{multicols}{2}
Only the sum of all diagrams is gauge invariant and is correspond to contact (or quark loop) contribution.

\section{LbL in AMM of muon}

The contribution of light by light process to AMM of muon has the form:
\begin{align}
&a_{\mu}^{\mathrm{LbL}}=\frac{e^{6}}{48m_{\mu}}\int\frac{d^{4}q_{1}}{(2\pi
	)^{4}}\int\frac{d^{4}q_{2}}{(2\pi)^{4}}\times\nonumber\\
&\quad\times\frac{{\Pi}_{\rho\mu\nu
		\lambda\sigma}(q_{2},-q_{3},q_{1})\mathrm{T}^{\rho\mu\nu\lambda\sigma}\left(
	q_{1},q_{2},p\right)  }{q_{1}^{2}q_{2}^{2}q_{3}^{2}((p+q_{1})^{2}-m_{\mu}%
	^{2})((p-q_{2})^{2}-m_{\mu}^{2})}, \label{aLbL2}%
\end{align}
where the tensor $\mathrm{T}^{\rho\mu\nu\lambda\sigma}$ is the Dirac trace
\begin{align}
&\mathrm{T}^{\rho\mu\nu\lambda\sigma}\left(  q_{1},q_{2},p\right)
=
\mathrm{Tr}\biggl(  (\hat{p}+m_{\mu})[\gamma^{\rho},\gamma^{\sigma}](\hat
{p}+m_{\mu})\times\nonumber\\
&\quad\quad\times\gamma^{\mu}(\hat{p}-\hat{q}_{2}+m_{\mu})\gamma^{\nu}(\hat{p}%
+\hat{q}_{1}+m_{\mu})\gamma^{\lambda}\biggr)  .\nonumber
\end{align}

Taking the Dirac trace, the tensor $\mathrm{T}^{\rho\mu\nu\lambda\sigma}$
becomes a polynomial in the momenta $p$, $q_{1}$, $q_{2}$.

After that, it is convenient to convert all momenta into the Euclidean space, and we will
use the capital letters $P$, $Q_{1}$, $Q_{2}$ for the corresponding
counterparts of the Minkowskian vectors $p$, $q_{1}$, $q_{2}$, e.g.
$P^{2}=-p^{2}=-m_{\mu}^{2}$, $Q_{1}^{2}=-q_{1}^{2}$, $Q_{2}^{2}=-q_{2}^{2}$.
Then Eq. (\ref{aLbL2}) becomes
\begin{align}
a_{\mu}^{\mathrm{LbL}}  &  =\frac{e^{6}}{48m_{\mu}}\int\frac{d_{E}^{4}Q_{1}%
}{(2\pi)^{4}}\int\frac{d_{E}^{4}Q_{2}}{(2\pi)^{4}}\frac{1}{Q_{1}^{2}Q_{2}%
^{2}Q_{3}^{2}}\frac{\mathrm{T}^{\rho\mu\nu\lambda\sigma}{\Pi}_{\rho
	\mu\nu\lambda\sigma}}{D_{1}D_{2}},\nonumber\\
&  D_{1}=(P+Q_{1})^{2}+m_{\mu}^{2}=2(P\cdot Q_{1})+Q_{1}^{2},\label{aLbL3}x\\
&  D_{2}=(P-Q_{2})^{2}+m_{\mu}^{2}=-2(P\cdot Q_{2})+Q_{2}^{2}.\nonumber
\end{align}

Since the highest order of the power of the muon momentum $P$ in
$\mathrm{T}^{\rho\mu\nu\lambda\sigma}$ is two \footnote{The possible
	combinations with momentum $P$ are
	\begin{align*}
	&  (P\cdot Q_{1})^{2}=(P\cdot Q_{1})(D_{1}-Q_{1}^{2})/2,\quad(P\cdot
	Q_{2})^{2}=-(P\cdot Q_{2})(D_{2}-Q_{2}^{2})/2,\\
	&  (P\cdot Q_{1})(P\cdot Q_{2})=-(D_{1}-Q_{1}^{2})(D_{2}-Q_{2}^{2})/4,\\
	\quad &  (P\cdot Q_{1})=(D_{1}-Q_{1}^{2})/2,\quad(P\cdot Q_{2})=-(D_{2}%
	-Q_{2}^{2})/2.
	\end{align*}
} and ${\Pi}_{\rho\mu\nu\lambda\sigma}$ is independent of \ $P$,
the factors in the integrand of (\ref{aLbL3}) can be rewritten as
\begin{align}
\frac{\mathrm{T}^{\rho\mu\nu\lambda\sigma}{\Pi}_{\rho\mu\nu
		\lambda\sigma}}{D_{1}D_{2}}=\sum\limits_{a=1}^{6}A_{a}\tilde{{\Pi}}%
_{a}, \label{TPbold}%
\end{align}
with the coefficients%
\begin{align}
&A_{1}=\frac{1}{D_{1}},\quad A_{2}=\frac{1}{D_{2}},\quad A_{3}=\frac{(P\cdot
	Q_{2})}{D_{1}},\quad A_{4}=\frac{(P\cdot Q_{1})}{D_{2}},\nonumber\\
&\quad A_{5}=\frac
{1}{D_{1}D_{2}},\quad A_{6}=1,
\end{align}
where all $P$-dependence is included in the $A_{a}$ factors, while
$\tilde{{\Pi}}_{a}$ are $P$-independent.

Then, one can average over the direction of the muon momentum $P$ (as was
suggested in \cite{Jegerlehner:2009ry} for the pion-exchange contribution)
\begin{align}
&\int\frac{d_{E}^{4}Q_{1}}{(2\pi)^{4}}\int\frac{d_{E}^{4}Q_{2}}{(2\pi)^{4}%
}\frac{A_{a}}{Q_{1}^{2}Q_{2}^{2}Q_{3}^{2}}...=\nonumber\\
&\quad\frac{1}{2\pi^{2}}%
\int\limits_{0}^{\infty}dQ_{1}\int\limits_{0}^{\infty}dQ_{2}\int%
\limits_{-1}^{1}dt\,\sqrt{1-t^{2}}\frac{Q_{1}Q_{2}}{Q_{3}^{2}}\langle
A_{a}\rangle...,\label{Av} %
\end{align}
where the radial variables of integration $Q_{1}\equiv\left\vert
Q_{1}\right\vert $ and $Q_{2}\equiv\left\vert Q_{2}\right\vert $ and the
angular variable $t=(Q_{1}\cdot Q_{2})/\left(  \left\vert Q_{1}\right\vert
\left\vert Q_{2}\right\vert \right)  $ are introduced.
The averaged $A_{a}$ factors was introduced in \cite{Jegerlehner:2009ry}

\begin{center}
	\includegraphics[width=0.45\textwidth]{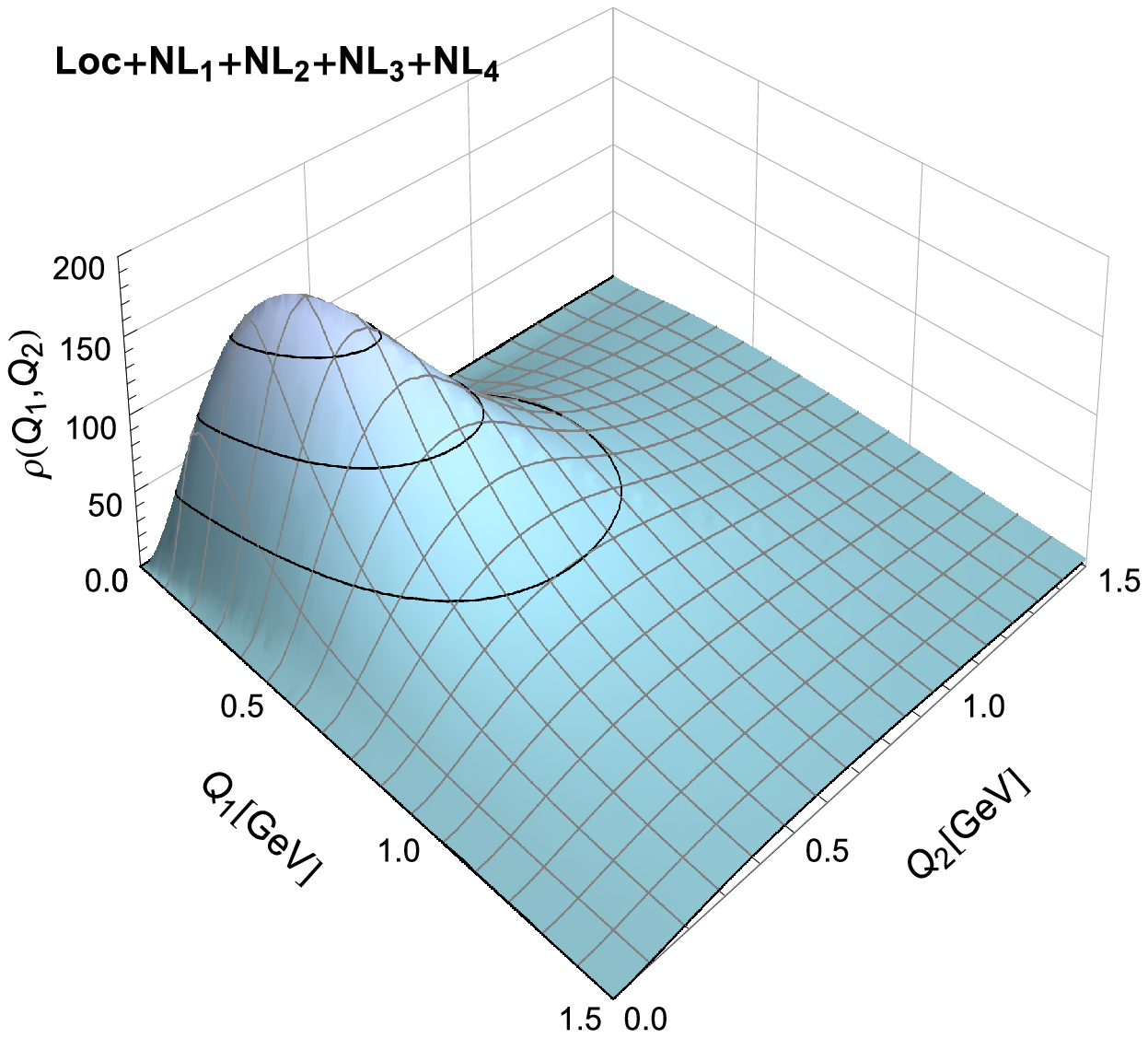} \vspace*{8pt}%
	\figcaption{The 3D density $\rho(Q_{1},Q_{2})$ defined in Eq. (\ref{dens}).}
	\label{Fig: 3Ddensity}%
\end{center}

\subsection{Density function}
For investigation of the dependence of contribution from photon legs virtuality 
one can watch for "density function". This is the function which corresponds to the LbL contribution  to AMM before integration over intermediate photons virtualities.
 \begin{equation}
\rho^{\mathrm{LbL}}(Q_{1},Q_{2})=\frac{Q_{1}Q_{2}}{2\pi^{2}}\sum
\limits_{a=1}^{6}\int\limits_{-1}^{1}dt\,\frac{\sqrt{1-t^{2}}}{Q_{3}^{2}%
}\langle A_{a}\rangle\tilde{{\Pi}}_{a}. \label{dens}%
\end{equation}

The volume under 3D -density function is full contribution to AMM of muon. In Fig.~\ref{Fig: Slice} this function is shown for nonlocal model in leading 1/$N_c$ order. 

One can see that contribution is mainly localized in a range of virtualities of photons around 1 GeV.

\begin{center}
	\includegraphics[height=60mm]{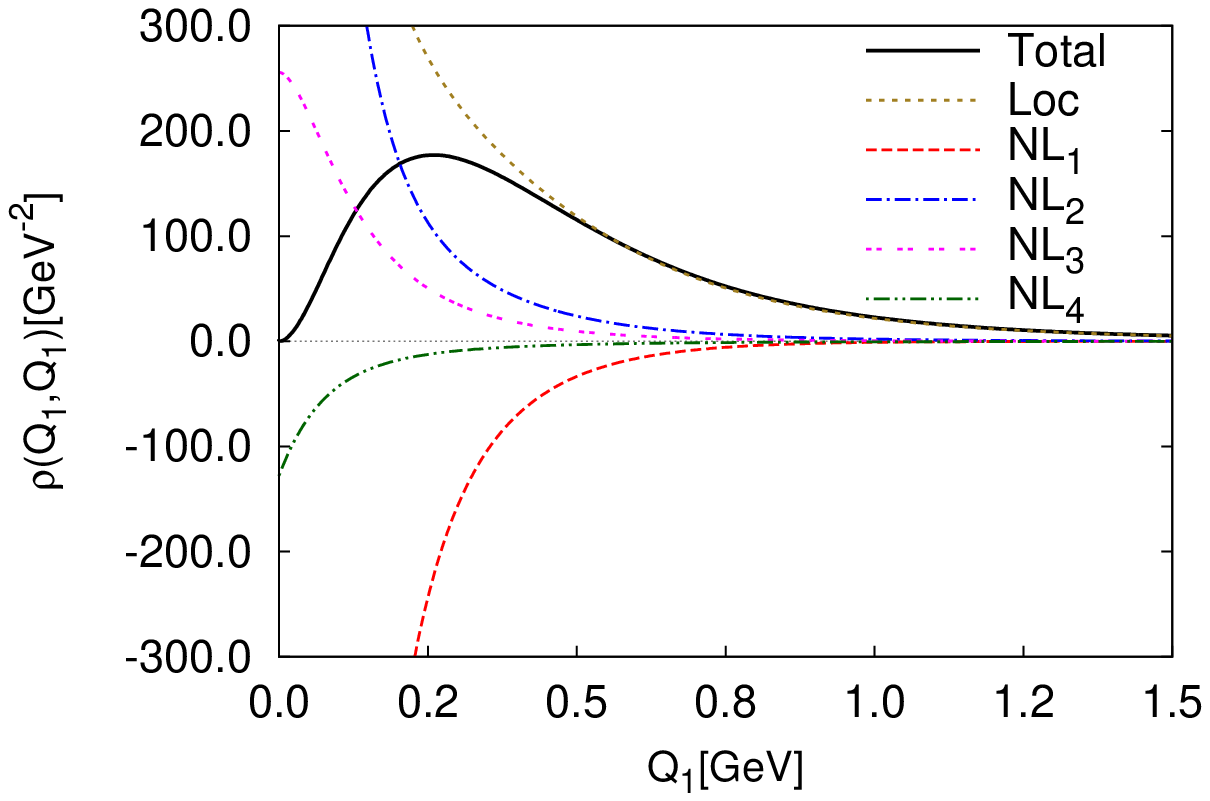}
	\figcaption{\label{Fig: Slice}The 2D slice of the density $\rho(Q_{1},Q_{2})$ at
		$Q_{2}=Q_{1}$. Different curves correspond to the contributions of
		topologically different sets of diagrams drawn in Fig. \ref{Fig: BoxCont}. The
		contribution of the box diagram with the local vertices, Fig.
		\ref{Fig: BoxCont}a, is the dot (olive) line(Loc); the box diagram, Fig.
		\ref{Fig: BoxCont}a, with the nonlocal parts of the vertices is the dash (red)
		line (NL$_{1}$); the triangle, Fig. \ref{Fig: BoxCont}b, and loop, Fig.
		\ref{Fig: BoxCont}c, diagrams with the two-photon vertices is the dash-dot
		(blue) line (NL$_{2}$); the loop with the three-photon vertex, Fig.
		\ref{Fig: BoxCont}d, is the dot-dot (magenta) line (NL$_{3}$); the loop with
		the four-photon vertex, Fig. \ref{Fig: BoxCont}e, is the dash-dot-dot (green)
		line (NL$_{4}$); the sum of all contributions (Total) is the solid (black)
		line. At zero all contributions are finite.
	}%
\end{center}
In Fig. \ref{Fig: Slice}, the slice of $\rho^{\mathrm{HLbL}}(Q_{1},Q_{2})$ in
the diagonal direction $Q_{2}=Q_{1}$ is presented together with the partial
contributions from the diagrams of different topology. One can see, that the
$\rho^{\mathrm{HLbL}}(0,0)=0$ is due to a nontrivial cancellation of different
diagrams of Fig. \ref{Fig: BoxCont}. This important result is a consequence of
gauge invariance and the spontaneous violation of the chiral symmetry, and
represents the low energy theorem analogous to the theorem for the Adler
function at zero momentum. Another interesting feature is, that the large
$Q_{1}$, $Q_{2}$ behavior is dominated by the box diagram with local vertices
and quark propagators with momentum-independent masses in accordance with perturbative theory.
All this is very important characteristics of the N$\chi$QM, interpolating the
well-known results of the chiral perturbative theory at low momenta and the
operator product expansion at large momenta. Earlier, similar results were
obtained for the two-point \cite{Dorokhov:2004ze} and
three-point \cite{Dorokhov:2005pg} correlators.

\section{Results and conclusion}

The contribution to AMM of muon from LbL process in N$\chi$QM correspond contributions from contact term and term with intermediate pseudoscalar and scalar channels. The contact term contribution is:
\begin{equation}
a_{\mu}^{\mathrm{HLbL,Loop}}=\left(  11.0\pm0.9\right)  \cdot10^{-10}.
\end{equation}
And the total contribution is estimated as 
\begin{equation}
a_{\mu}^{\mathrm{HLbL}}= 16.8 (1.25) \cdot10^{-10},
\end{equation}
where error bar is the band in region of physical dynamical mass.

In Fig.~\ref{ContAll} one can see that it is important, at least in the framework of quark model, to take into account not only diagrams with intermediate mesons but also the contact term (or quark loop) contribution.

In comparison with other model calculation, our results are quite close to the recent results obtained in
\cite{Fischer:2010iz,Greynat:2012ww}. 
The specific feature of our model and Dyson-Schwinger approach \cite{Fischer:2010iz} is that the
due to nonlocal interaction kernel the quarks becomes dynamical one with momentum-dependent mass.
The predictions of the N$\chi$QM for the different contributions to the muon $g-2$ are in agreement with \cite{Fischer:2010iz} within $10\%$.

The next step of calculations is to extend quark model in order to estimate subleading in 1/$N_c$ terms. This subleading contribution from diagrams with meson loop has negative sign 
\cite{Kinoshita:1984it}. 
\begin{center}
	\includegraphics[height=60mm]{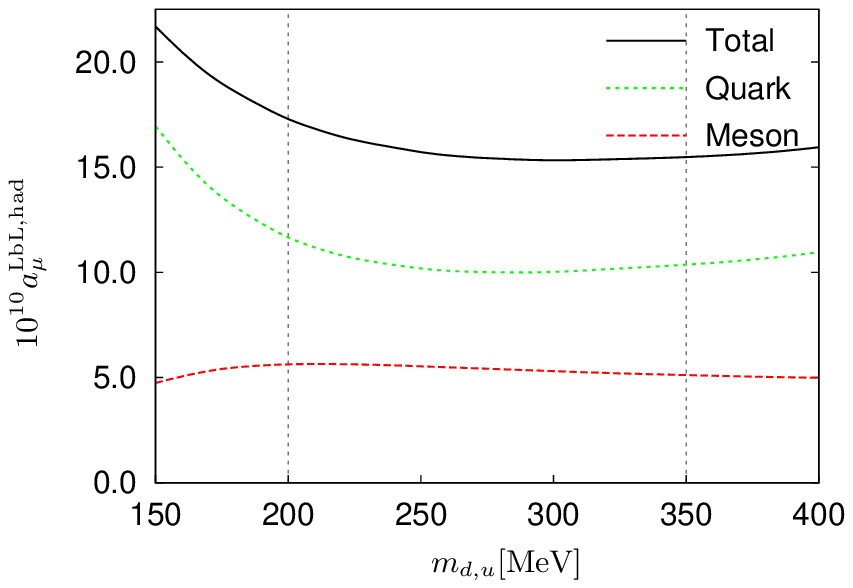}
		\figcaption{\label{ContAll} The contribution of muon AMM from LbL depending on the dynamical mass of quark $m_{d,u}$ in zero order on 1/$N_c$. The solid curve is total contribution, red and green dashed curve --- contact terms Fig.~\ref{Fig: BoxCont} and meson exchange \cite{Dorokhov:2011zf,Dorokhov:2012qa} respectively.}
\end{center}

For solving the puzzle of LbL contribution, we should to better understanding the physics of strong interaction at long distance. In principle one can do this with more accurate measurement of meson form factors.

\end{multicols}
\vspace{-1mm}
\centerline{\rule{80mm}{0.1pt}}
\vspace{2mm}

\begin{multicols}{2}

\end{multicols}

\clearpage

\end{CJK*}
\end{document}